\documentclass[conference, a4paper]{IEEEtran}
\usepackage{cite}
\usepackage[latin1]{inputenc}
\usepackage[english]{babel}
\usepackage{amsmath,amsfonts,amssymb}
\usepackage{graphicx, epsfig}
\usepackage{subfigure}
\usepackage{array}
\usepackage{multirow}
\usepackage{mdwmath}
\usepackage{mdwtab}
\usepackage{algorithmic}
\usepackage{algorithm}
\usepackage{color}

\ifCLASSINFOpdf

\else

\fi

\begin{document}
\title{\huge Iterative Detection and LDPC Decoding Algorithms for MIMO Systems in Block-Fading Channels}

\author{\IEEEauthorblockN{Andre G. D. Uchoa, Peng Li, Rodrigo C. de Lamare and Cornelius Healy}
\IEEEauthorblockA{\\University of York,
Heslington York, YO10 5DD, UK\\
$\lbrace$agdu500, pl534, rcdl500, cth503$\rbrace$@york.ac.uk}}

\maketitle

\begin{abstract}
We propose an Iterative Detection and Decoding (IDD) scheme with Low Density Parity Check (LDPC) codes for Multiple Input Multiple Output (MIMO) systems for block-fading $F = 2$ and fast fading Rayleigh channels. An IDD receiver with soft information processing that exploits the code structure and the behaviour of the log likelihood ratios (LLR)'s is developed. Minimum Mean Square Error (MMSE) with Successive Interference Cancellation (SIC) and with Parallel Interference Cancellation (PIC) schemes are considered. The soft \textit{a posteriori} output of the decoder in a block-fading channel with Root-Check LDPC codes has allowed us to create a new strategy to improve the Bit Error Rate (BER) of a MIMO IDD scheme. Our proposed strategy in some scenarios has resulted in up to 3dB of gain in terms of BER for block-fading channels and up to 1dB in fast fading channels.
\end{abstract}


% \fi
\IEEEpeerreviewmaketitle

\section{Introduction}
%Motivation
The most recent IEEE Wireless Local Area Network (WLAN) 802.11ad standard suggests that to achieve high throughput the devices must operate with LDPC codes. Since a WLAN MIMO system is subject to multi-path propagation and mobility, this wireless system is
characterized by time-varying channels with fluctuating signal strength. In applications subject to delay constraints and slowly-varying channels, only limited independent fading
realizations are experienced. In such conditions also known as non-ergodic scenarios, the channel capacity is zero since there is an irreducible probability, termed outage probability \cite{rasmussen.10}, that the transmitted data rate is not supported
by the channel. A simple and useful model that captures the essential characteristics of non-ergodic channels is the block-fading channel \cite{rappaport}. It is especially important in wireless communications with slow time-frequency hopping (e.g., cellular networks and wireless local area networks) or multi-carrier modulation using Orthogonal Frequency Division Multiplexing (OFDM) \cite{boutros.07}. Codes designed for block-fading channels are expected to achieve the channel diversity and to offer good coding gains.

A family of LDPC codes called Root-Check for block-fading channels was proposed in \cite{boutros.07}. Root-check codes are able to achieve the maximum diversity of a block-fading channel and have a performance near the limit of outage when decoded using the
Sum Product Algorithm (SPA). Root-check codes are always designed with code rate $R = 1/F$, since the Singleton bound determines that this is the highest code rate possible to obtain the maximum diversity order \cite{boutros.07}. Several works have been proposed to improve the coding gain, and to address problems such as low complexity encoding and even low memory storage. The most recent works in the literature in terms of Root-Check construction are found in \cite{salehi.10, iswcs.11, peg.comms.11, iswcs.12, iswcs.13}.

%Transiction from Root-Check to MIMO must be less abrupt
Spatially multiplexed coded modulations that are decoded using Iterative Detection and Decoding (IDD) techniques can offer capacity-approaching performance. One seminal work that introduced the concept of IDD schemes was reported in \cite{poor.99}. In \cite{poor.99}, it is proposed a low-complexity iterative multiuser receivers for coded code-division multiple-access (CDMA) systems over multipath channels. It is shown that as the number of outer iteration (the message exchanged between detector and decoder) increase their scheme is capable to get close to the single user capacity. The work done in \cite{pingli.05} is essentially an extension of \cite{poor.99} applied for Multiple Input Multiple Output (MIMO) systems even though, it is suggested that scheme in \cite{pingli.05} is computationally more efficient than \cite{poor.99} and it can be applied to multi-user detection.

In \cite{lamare.tcomms.08}, the authors proposed an MMSE decision feedback (DF) detector that employs a successive parallel arbitrated DF structure based on the generation of parallel arbitrated branches. The motivation for that was to mitigate the effects of error propagation often found in interference cancellation structures. The results in \cite{lamare.tcomms.08} showed for both uncoded and convolutionally encoded systems using Viterbi and turbo decoding that the detection schemes proposed can offer considerable gains as compared to existing DF and linear receivers. In \cite{matsumoto.03} it is proposed a cancellation technique to mitigate Inter Symbol Interference (ISI) and Multiple Access Interference (MAI) for frequency-selective MIMO channels. Moreover, the authors in \cite{matsumoto.03} have shown a scheme for channel estimation which is more realistic in terms of practical systems.

The work done in \cite{rfa.iet.11} proposes a SIC strategy for MIMO spatial multiplexing systems based on a structure with multiple interference cancellation branches. This multi-branch SIC (MB-SIC) structure employs multiple SIC schemes in parallel and each branch detects the signal according to its respective ordering pattern. By selecting the branch which yields  the estimates with the best performance according to the selection rule, the MB-SIC detector, therefore, achieves higher detection diversity. A low-complexity multiple feedback successive interference cancellation (MF-SIC) is proposed in \cite{peng.twc.11} for the uplink of multiuser MIMO (MU-MIMO) systems. Also, the authors have made an extension of the work in \cite{rfa.iet.11} to combine the MF-SIC with MB-SIC to achieve a higher detection diversity order. The results presented in \cite{peng.twc.11} show that their schemes significantly outperform the conventional SIC scheme and approach the optimal detector.

All the works previously mentioned deal with quasi-static Rayleigh fading channels or even fast Rayleigh fading channel. However, there is very little research in the literature related to the case of block-fading channels with MIMO systems. To the best of our knowledge, the only work which discusses the MIMO under block-fading channels is \cite{mimobf.11}. In \cite{mimobf.11}, the authors claimed that using Root-Check LDPC codes the MIMO system can achieve the desired channel diversity. Furthermore, they presented a relationship between the maximum LDPC code rate and $n_{tx}$ transmitting antennas is $R = \frac{1}{n_{tx}}$. In \cite{mimobf.11}, it is only considered the direct application of Root-Check LDPC codes for MIMO in block-fading channels. In contrast, in our work we discuss how we can improve the system performance based on the log likelihood ratio (LLR) behaviour of a Root-Check LDPC code. We have observed that if we properly manipulate the LLR output of the decoder and exploit the code structure we can obtain gains of up to $3dB$ with respect to non Root-Check LDPC codes in a IDD system. Moreover, we have considered several scenarios of MIMO signalling. The main contribution of our work is to develop a novel IDD scheme that exploits the code structure and provides a new strategy for LLR manipulation that improves the performance of MIMO IDD systems under block-fading channels. Furthermore, the improved performance noticed in block-fading channels instigated us to consider the case of fast-fading channels which also results in a better performance than traditional IDD schemes. The most important aspect to be considered is that the gains provided by our proposed IDD scheme do not require extra computationally effort or even any extra memory storage.

%How the paper is organized
The rest of this paper is organized as follows. In Section II we describe the system model. In section III we discuss our proposed LLR strategy. Section IV it is depicted the simulation results, while Section V concludes the paper.

\section{System Model}
\begin{figure}[htb]
 \centering
\resizebox{88mm}{!}{
\includegraphics{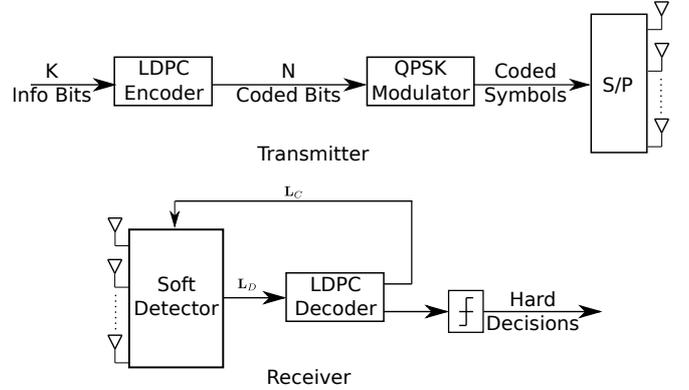}
}\caption{Transmitter and receiver structure of an LDPC coded MIMO system. The receiver is based on iterative detection-and-decoding scheme.} \label{fig:fig1}
\end{figure}

Consider a Root-Check LDPC-coded MIMO system with $n_{tx}$ transmitter antennas and $n_{rx}$ receiver antennas, signalling through a block fading channel, where $F$ is the number of independent fading blocks per codeword of length $N$. The transmitter and receiver structure are illustrated in Fig. \ref{fig:fig1}. The data is first encoded by a Root-Check LDPC code with rate $\frac{1}{F}$, modulated by a complex constellation with $2^{M_{c}}$ possible signal points and average energy equal to $E_{s}$, and then distributed among the $n_{tx}$ antennas. Let $\mathbf{s}_{t}$ be an $n_{tx} \times 1$ vector of transmitted symbols and $\mathbf{r}_{t}$ an $n_{rx} \times 1$ vector of received signals related by:
\begin{equation} \label{eq:recsymbol}
 \mathbf{r}_{t} = \mathbf{H}_{f}\mathbf{s}_{t}+\mathbf{n}_{g_{t}},
\end{equation}
where $t=\{1,2,\cdots, 2^{M_{c}}\}$, $f=\{1,2,\cdots,F\}$, $f$ and $t$ are
related by $f=\lceil F\frac{t}{2^{M_{c}}}\rceil$, where $\lceil \phi \rceil$
returns the smallest integer not smaller than $\phi$, $\mathbf{H}_f$ is the
$n_{rx}~\times~n_{tx}$ real complex channel matrix with Rayleigh fading coefficient of the $f$-th block and $\mathbf{n}_{g_{t}}$ is a vector of independent zero-mean, complex Gaussian noise entries with variance $\sigma^{2} = N_0/2$ per real component. In the case of fast fading we assume that each received symbol $\mathbf{r}_{t}$ will be under a distinct fading coefficient, which means $F = N$. Also we consider that the average signal-to-noise ratio (SNR) at each receiver antenna is independent of the number of transmitter antennas $n_{tx}$. So, the SNR can be defined as $n_{tx}\cdot \frac{E_{s}}{N_{0}}$. In this paper, we assume that the transmitted symbols $\mathbf{s}_{t}$ are taken from a modulation constellation $\mathcal{A} = \lbrace a_{1}, a_{2}, \cdots, a_{C}\rbrace$. The information transmission rate is $R=K/N$, where $K$ is the number of information bits per codeword of length $N$. For the case of a block-fading channel, we consider $R=\frac{1}{F}=\frac{1}{n_{tx}}$, since then it is possible to design a practical diversity achieving code \cite{salehi.10}.

An IDD scheme is used to approach the maximum-likelihood (ML) performance of joint MIMO detection and LDPC decoding. Fig. \ref{fig:fig1} gives a block diagram of the turbo iterative receiver structure. In this structure, the soft MIMO detector incorporates extrinsic information provided by the LDPC decoder, and the LDPC decoder incorporates soft information provided by the MIMO detector. Extrinsic information between the detector and decoder is then exchanged in an iterative fashion until an LDPC codeword is found or a maximum number of iterations is performed. We will call inner iterations the iterations done by the log Sum Product (SP) LDPC decoder, and outer iterations the iterations between the decoder and the detector.

%The performance of a MIMO communication system in a non-ergodic block-fading channel can
%be investigated by means of the outage probability \cite{telatar.99}, which is defined as:
%\begin{equation} \label{eq:pout}
%P_{out} = {\cal P}(\cal I < \cal R ),
%\end{equation}where ${\cal P}(\phi)$ is the probability of event $\phi$ and $\cal I$ is the instantaneous mutual information after channel realization. The MIMO mutual information $I_{G}(SNR,\mathbf{H})$, for Gaussian channel inputs is\cite{salehi.10}:
%\begin{equation}\label{eq:outage}
%I_{G}(SNR,\mathbf{H}_{f})=\frac{1}{\vert \mathcal{A} \vert} \cdot \log_2 \det(\mathbf{I}_{n_{rx}} + \frac{SNR}{n_{tx}}\cdot \mathbf{H}_{f}\mathbf{H}_{f}^{H}),
%\end{equation} where $\mathbf{I}_{n_{rx}}$ is an identity matrix of size $n_{rx}\times n_{rx}$ and $(\cdot)^{H}$ stands for the Hermitian transpose operation. Wherefore, an outage occurs when the average accumulated mutual information among blocks is smaller than the attempted information transmission rate.

\subsection{MAP Detector}
In the soft maximum a posteriori (MAP) detector, the received vector $\mathbf{r}_{t}$ is demapped by LLR calculation for each bit of the $n_{tx}M_{c}$ coded bits included in the transmit vector $\mathbf{s}_{t}$. The \textit{extrinsic} information provided by the MAP detector is the difference of the soft-input and soft-output LLR values on the coded bits. For the t$-th$ code bit $b_{t}$ ($i~\in \lbrace 1,\cdots, n_{t}M_{c}\rbrace$) of the transmit vector $\mathbf{s}_{t}$, the extrinsic LLR value of the estimated bit is computed as
{\small
\begin{eqnarray} \label{eq:ext_bit}
L_{D}(b_{t}) & = & \log \frac{P(b_{t} = +1)\vert r_{t}, \mathbf{H}_{f}}{P(b_{t} = -1)\vert r_{t}, \mathbf{H}_{f}}~-~\log \frac{P(b_{t} = +1)}{P(b_{t} = -1)} \nonumber \\
& = & \log \frac{\sum_{\mathbf{s}_{t}~\in \chi_{t}^{+1}}P(\mathbf{r}_{t}\vert s_{t}, \mathbf{H}_{f})P(\mathbf{s}_{t})}{\sum_{\mathbf{s}_{t}~\in \chi_{t}^{-1}}P(\mathbf{r}_{t}\vert s_{t}, \mathbf{H}_{f})P(\mathbf{s}_{t})}~-~L_{C}(b_{t}),
\end{eqnarray}}
where $L_{C}(b_{t})$ is the extrinsic information of the bit $b_{t}$ computed by the LDPC decoder in the previous turbo iteration ($L_{C}(b_{t}) = 0$ at the first iteration) and $\chi_{t}^{+1}$ is the set of $2^{n_{tx}M_{c}-1}$ vector hypotheses $\mathbf{s}_{t}$ having $b_{t} = +1$, ($\chi_{t}^{-1}$ is similarly defined). Assuming the bits within $\mathbf{s}_{t}$ are statistically independent of one another, the \textit{a priori} probability $P(\mathbf{s}_{t})$ can be written as
\begin{equation} \label{eq:apriori_prob}
P(\mathbf{s}_{t}) = \prod_{j = 1}^{n_{tx}M_{c}} P(b_{t}) = \prod_{j = 1}^{n_{tx}M_{c}} \left[1+exp(-\mathbf{s}_{t}^{b_{j}}L_{C}(b_{j})) \right]^{-1},
\end{equation}
where $\mathbf{s}_{t}^{b_{j}}$ corresponds to the value (+1,-1) of the j$-th$ bit in the vector $\mathbf{s}_{t}$. In the above LLR value calculation, the likelihood function $P(\mathbf{r}_{t}\vert s_{t}, \mathbf{H}_{f})$ is especified by a multivariate Gaussian density function.

\subsection{Detection Techniques for Spatial Multiplexing}
The optimal detection algorithm is the maximum likelihood (ML) detection algorithm given by
\begin{equation} \label{eq:ml_dect}
\mathbf{\hat{S}}_{{ML}_t} = \underset{\hat{S}\in \mathit{A}}{\operatorname{arg~min}}\Vert \mathbf{r}_{t}~-~\mathbf{H}_{f}\mathbf{\hat{S}}_{t} \Vert^{2},
\end{equation}
where $A$ denotes a set of $n_{tx}$-dimensional candidate vectors. The computational complexity, which increases exponentially with the number of transmit antennas, prevents the practical application of the ML detector. The MMSE linear detector is a relatively simple strategy to separate the transmitted signals at the receiver. It corresponds to designing an $n_{tx}\times n_{rx}$ parameter matrix $\mathbf{W}$ according to the MMSE criterion. The design of the MMSE filter matrix $\mathbf{W}$ is based on the minimization of the following cost function
\begin{equation} \label{eq:mmse_cost_func}
J(\mathbf{W}) = E\left[\Vert \mathbf{s}_{t} - \mathbf{W}^{H} \mathbf{r}_{t}\Vert^{2} \right],
\end{equation}
where $E[\cdot]$ stands for the expected value and $(\cdot)^{H}$ is the Hermitian transpose operation. By computing the gradient of (\ref{eq:mmse_cost_func}) with respect to $\mathbf{W}$ and then equating it to a null matrix, we obtain the $n_{rx} \times n_{tx}$ MMSE filter matrix
\begin{equation} \label{eq:mmse_filter_matrix}
\mathbf{W} = \left(\mathbf{H}\mathbf{H}^{H} + \frac{\sigma_{v}^{2}}{\sigma_{s}^{2}} \mathbf{I} \right)^{-1} \mathbf{H},
\end{equation}
where $\sigma_{v}^{2}$ is the noise variance, $\sigma_{s}^{2}$ is the signal power and $\mathbf{I}$ is the identity matrix. The MMSE linear detector expression (\ref{eq:mmse_filter_matrix}) requires the channel matrix $\mathbf{H}$ (in practice an estimate of it) and the noise variance $\sigma_{v}^{2}$ at the receiver. There are a number of other strategies to achieve the capacity gain of MIMO systems in which the V-BLAST \cite{vblast} is the most competitive one because of its lower complexity and good performance. Nevertheless, there is still a large performance gap between linear algorithms and ML-type detectors.

The SIC algorithm is usually combined with MMSE filtering, resulting in V-BLAST \cite{vblast} receivers. This provides improved performance at the cost of increased computational complexity. Rather than jointly decoding the transmitted signals, this nonlinear detection scheme first detects the first row of the signal and then cancels its effect from the overall received signal vector. It then proceeds to the next row. The reduced channel matrix now has dimension $n_{rx} \times (n_{tx}-1)$ and the signal vector has dimension $(n_{tx}-1)\times 1$. It then does the same operation on the next row. The channel matrix now reduces to $n_{rx} \times (n_{tx}-2)$ and the signal vector reduces to $(n_{tx}-2)\times 1$ and so on. If we assume that all the decisions at each layer are correct, then there is no error propagation. Otherwise, the error rate performance is dominated by the weakest stream, which is the first stream decoded by the receiver. Hence, the improved diversity performance of the succeeding layers does not help. To get around this problem the ordered SIC receiver was introduced \cite{stcmimo.book}. In this case, the signal with the strongest signal-to-interference-noise (SINR) ratio is selected for processing. This improves the quality of the decision and reduces the chances of error propagation. This is like an inherent form of selection diversity wherein the signal with the strongest SNR is selected \cite{stcmimo.book}.

%The ordered SIC algorithm is as follows:
%\begin{itemize}
%\item \textit{Ordering}: Determine the optimal detection order by choosing the row with
%minimum Euclidian norm (strongest SINR);
%\item \textit{Nulling}: Estimate the strongest transmit signal by nulling out all the weaker transmit signals;
%\item \textit{Slicing}: Detect the value of the strongest transmit signal by slicing to the nearest signal constellation value;
%\item \textit{Cancellation}: Cancel the effect of the detected signal from the received signal vector to reduce the detection complexity for the remaining signals.
%\end{itemize}
%
%For the case of the parallel interference cancellation (PIC) has the following steps:
%\begin{itemize}
%\item \textit{Select}: Get the first row of the channel matrix;
%\item \textit{Nulling}: Start nulling out the first signal from all the transmitted signals;
%\item \textit{Slicing}: Detect the value of the strongest transmit signal by slicing to the nearest signal constellation value;
%\item \textit{Cancellation}: Cancel the effect of the detected signal from the received signal vector to reduce the detection complexity for the remaining signals.
%\end{itemize}

\section{Proposed IDD Scheme and LLR Strategy}
\begin{figure}[htb]
\subfigure[][]{
  \centering
\includegraphics[width=41mm]{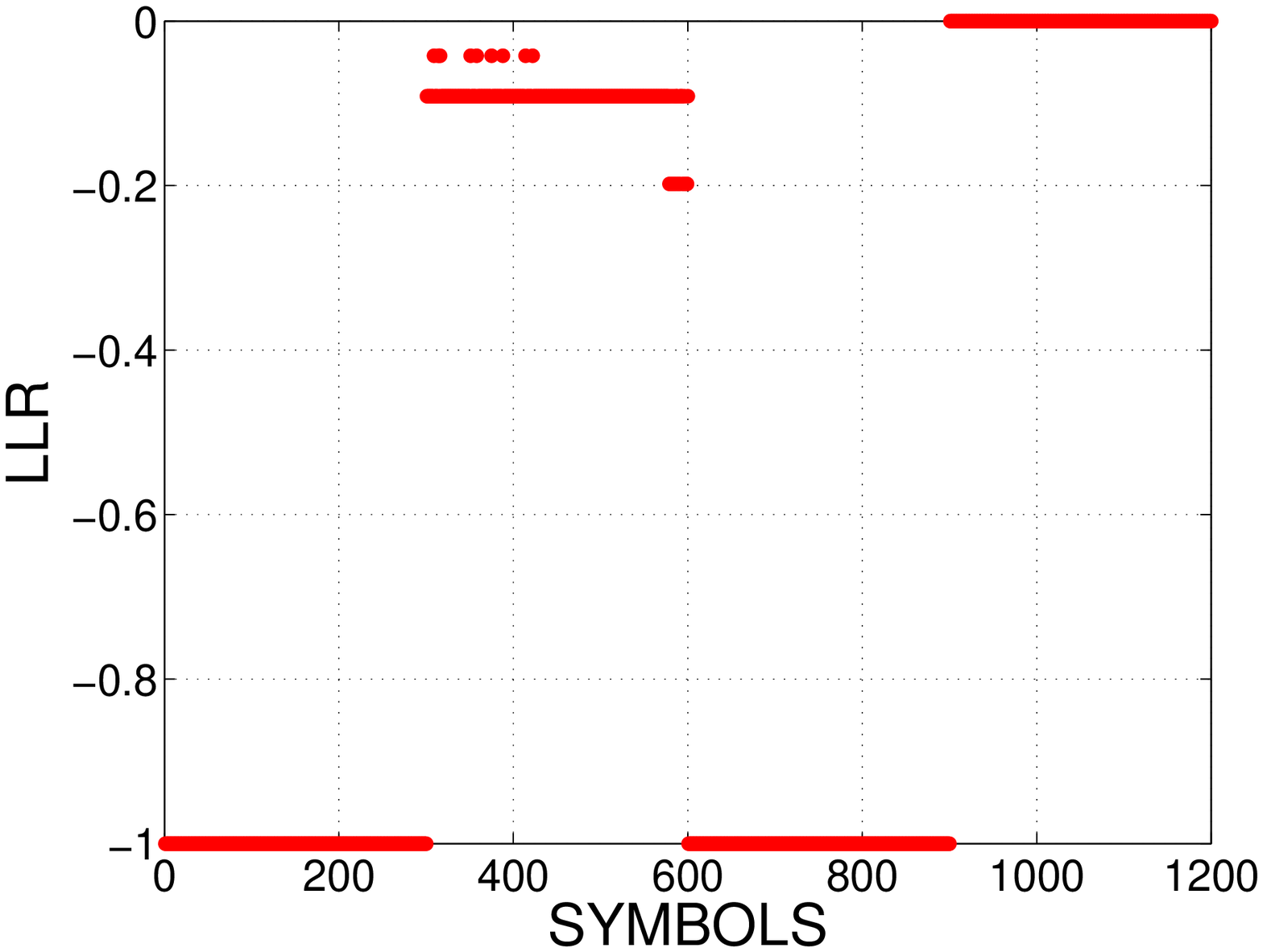}
\label{fig:llr_out_bpsk}
}
\subfigure[][]{
  \centering
\includegraphics[width=41mm]{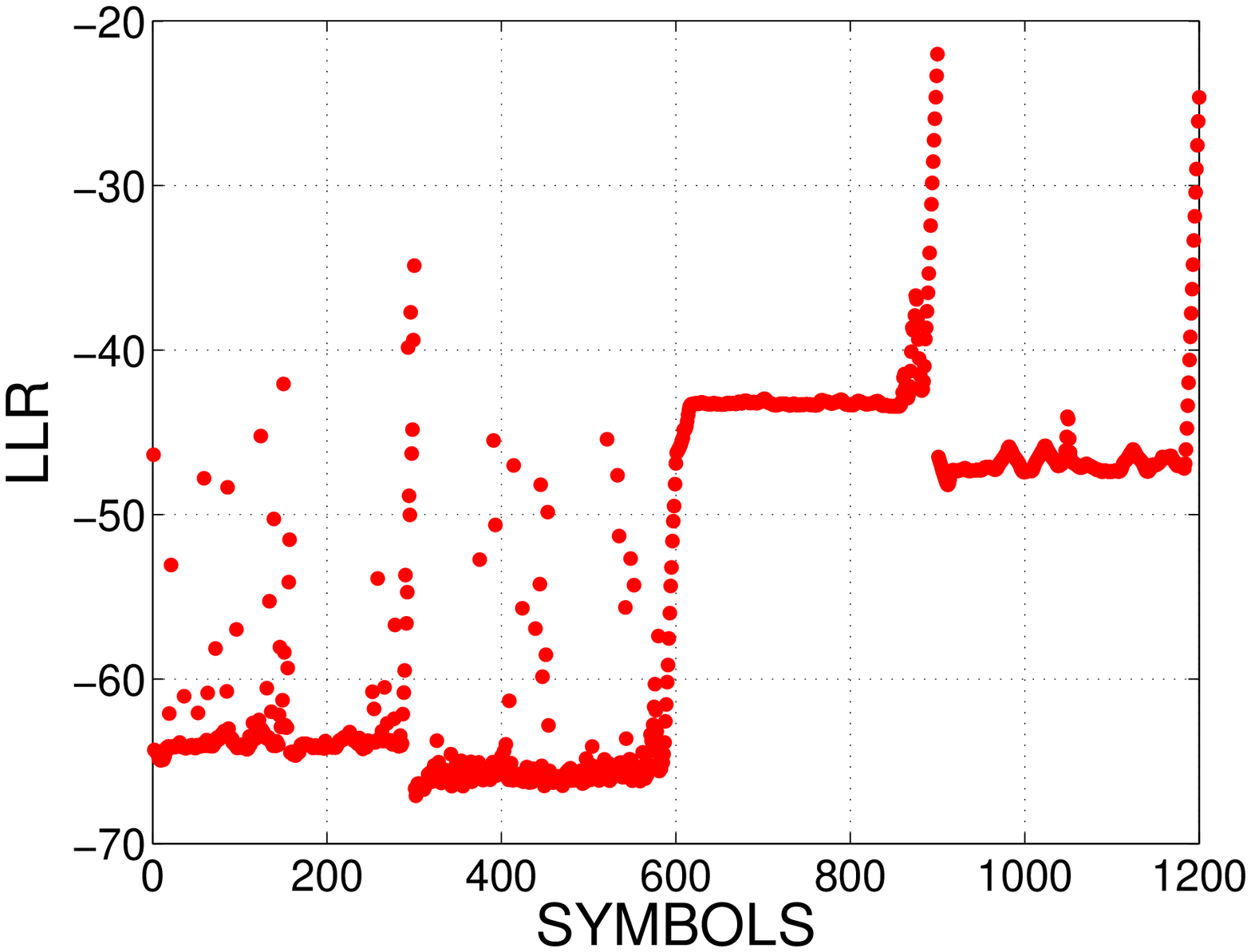}
\label{fig:llr_out_dec}
}

\subfigure[][]{
  \centering
\includegraphics[width=41mm]{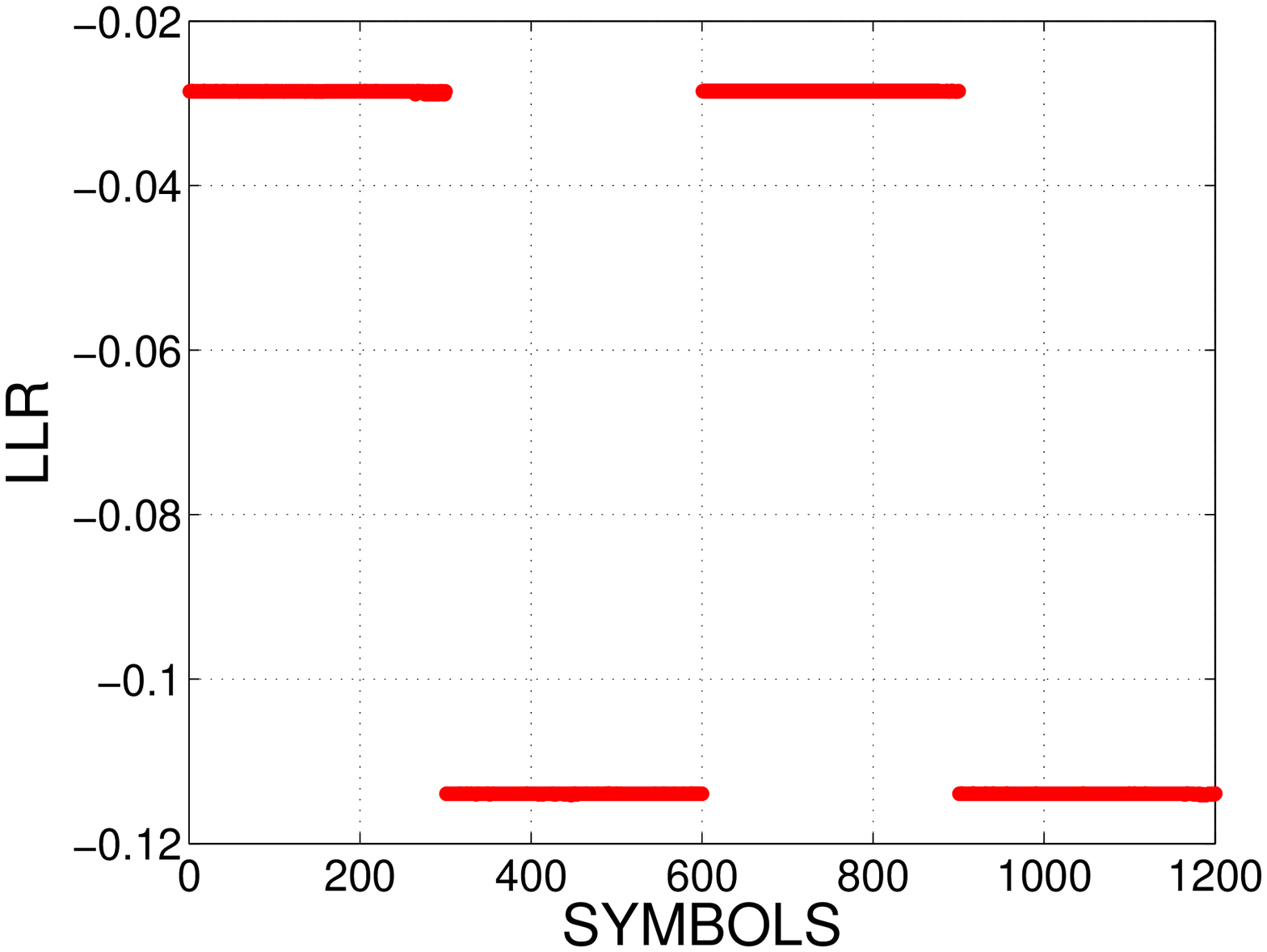}
\label{fig:llr_out_dec_opt}
}
\subfigure[][]{
  \centering
\includegraphics[width=41mm]{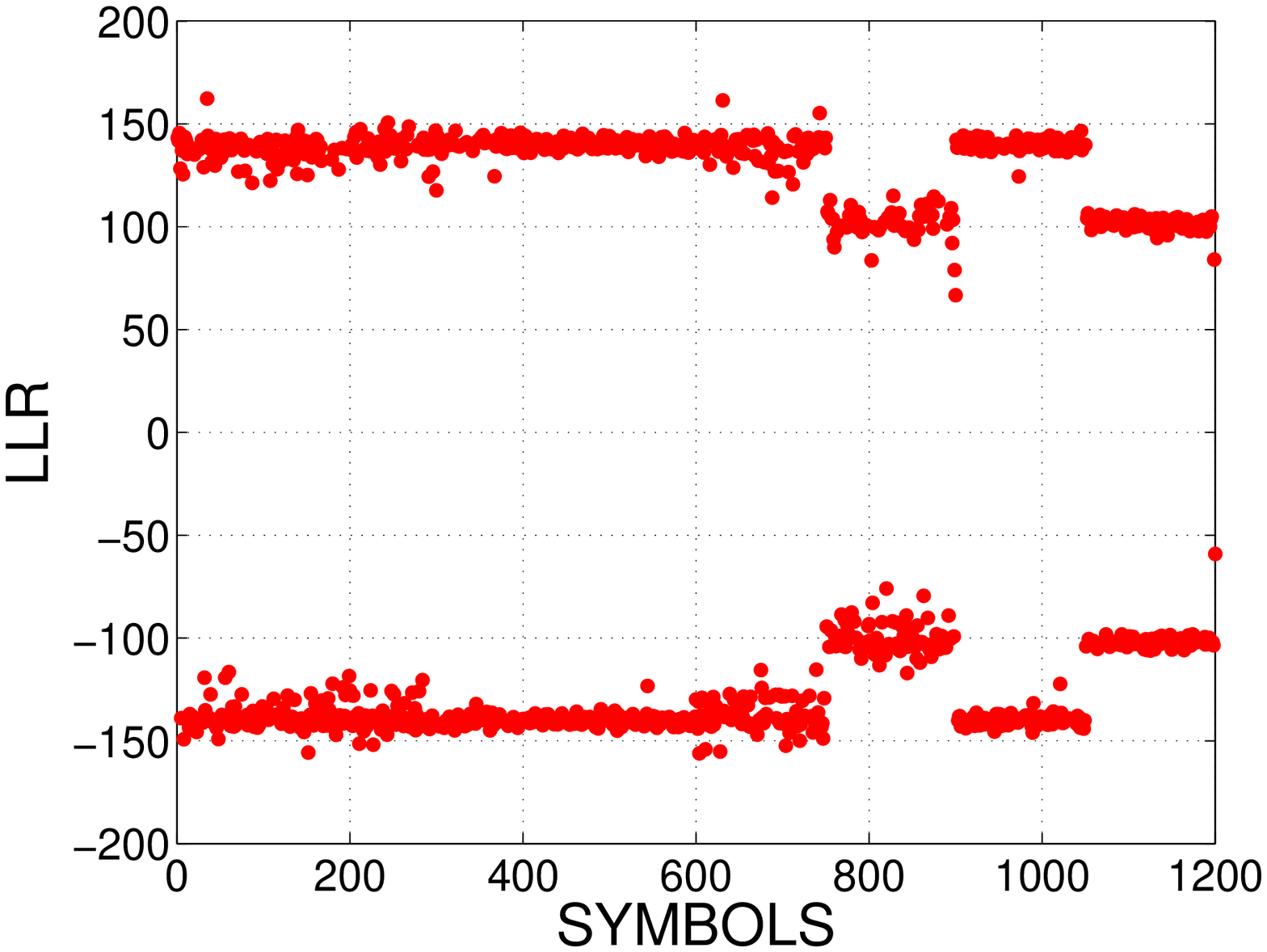}
\label{fig:llr_output_non_comp}
}
\subfigure[][]{
  \centering
\includegraphics[width=41mm]{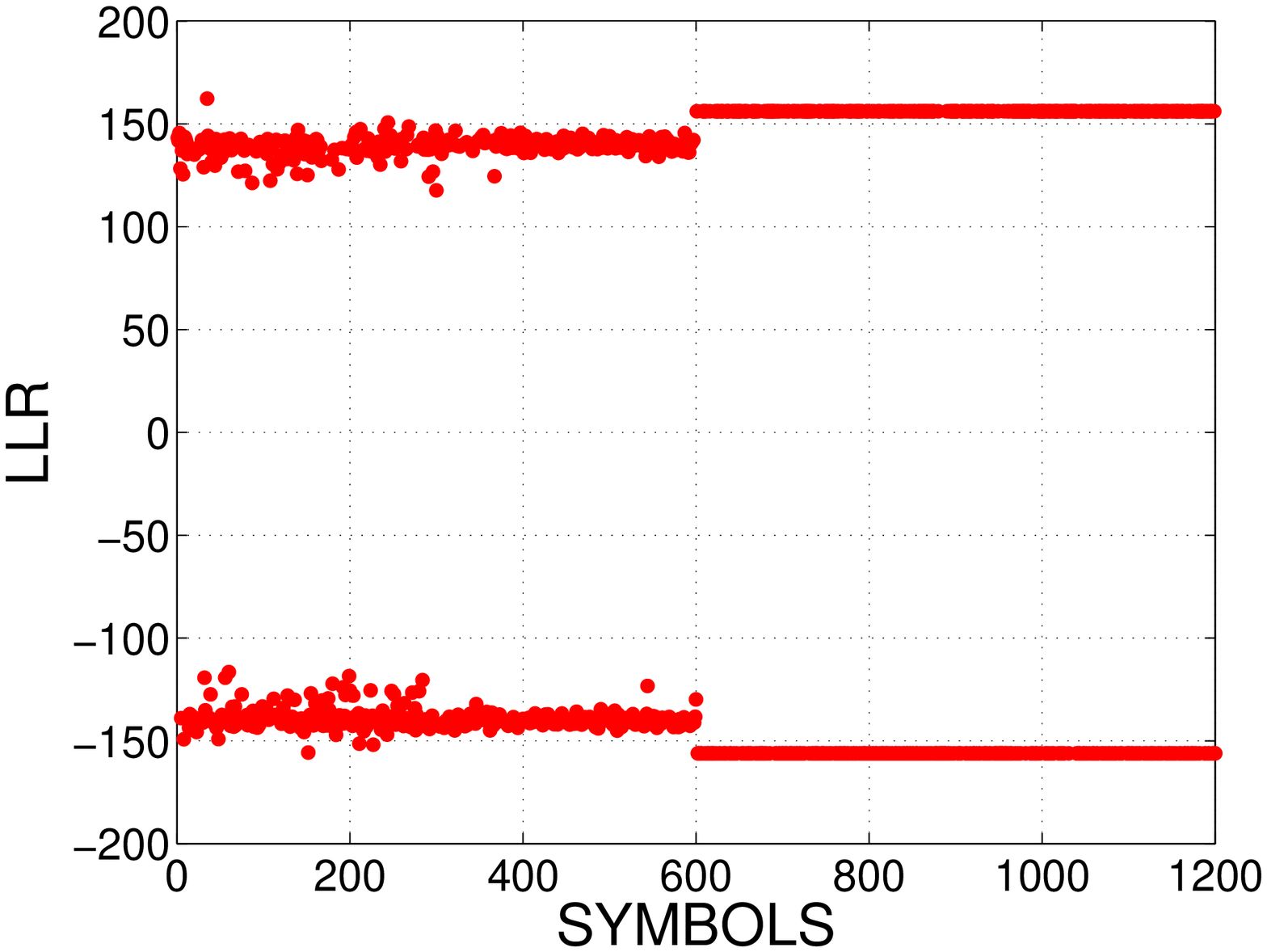}
\label{fig:llr_output_comp}
}
\subfigure[][]{
  \centering
\includegraphics[width=41mm]{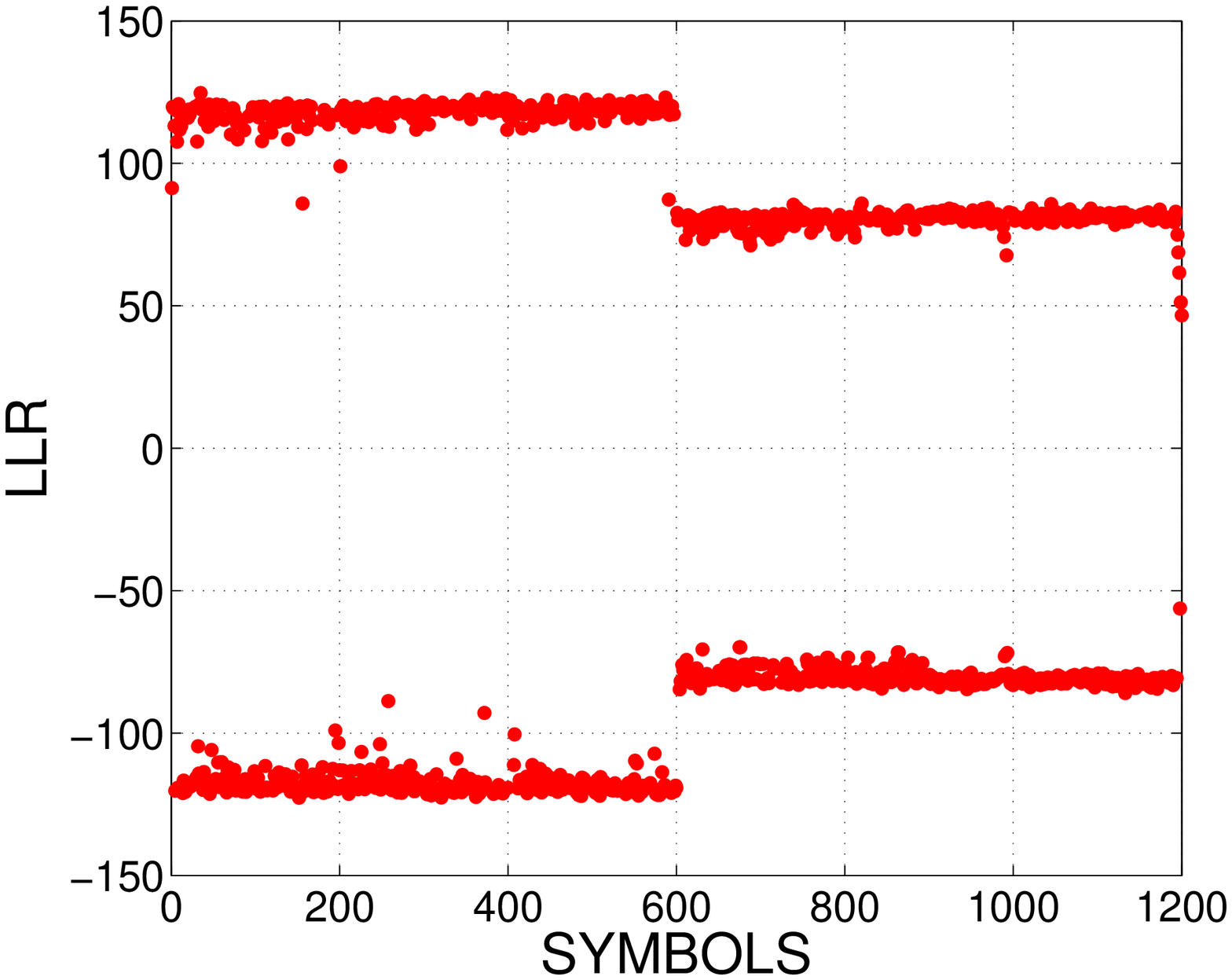}
\label{fig:llr_out_non_comp_ira} }
\caption{ LLR outputs of
Root-Check LDPC codes over a block-fading channel with $F = 2$ and
$S = 1200$ symbols. The maximum number of iterations is 20. In
\subref{fig:llr_out_bpsk} is the LLR output of all null codeword;
\subref{fig:llr_out_dec} describes the LLR output over a complex
block-fading channel; \subref{fig:llr_out_dec_opt} shows the LLR
output assuming the channel envelope;
\subref{fig:llr_output_non_comp} the LLR output of a Doping
Root-Check LDPC code assuming the channel envelope;
\subref{fig:llr_output_comp} applying the proposed scheme LLR-SP-BF
with respect to \subref{fig:llr_output_non_comp};
\subref{fig:llr_out_non_comp_ira} LLR output of an IRA LDPC code
assuming the channel envelope.} \label{fig:llr_analisys}
\end{figure}

We started by investigating the performance of the Root-Check LDPC codes under MIMO systems with IDD schemes. Several simulations were conducted under different scenarios which lead to similar results, where Root-Check LDPC codes lose in terms of bit error rate (BER) to the standard LDPC codes at high SNR. In addition, Fig. \ref{fig:llr_analisys} will provide a set of sub-plots which are important to understand how our proposed IDD scheme of LLR strategy are devised.

Let us consider a simple scenario where the Root-Check LDPC codes are expected to obtain full diversity in a single input single output (SISO) system. An arbitrary modulation was used, $F = 2$ fadings per codeword, code rate $\frac{1}{F} = \frac{1}{2}$, channel fading coefficients $h_{1}$ and $h_{2}$, all null code word will be sent through the "channel". For practical purposes, let us consider $h_{1} = 1$ and $h_{2} = 0$. Furthermore, we assume that there is no noise contribution from the receiver side for this particular experiment. As a result, we will be analysing only the influence of the fading under the soft decision at the decoder. The received code word will be,
\begin{equation} \label{eq:rcv_cw}
y_{t} = s_{t}\cdot h_{f},
\end{equation}
where $1 \leq t \leq N$, $y_{t}$ leads to something like,
\begin{equation} \label{eq:rcv_mod}
\mathbf{Y} = -1\cdot h_{1}, \cdots, 0\cdot h_{2}, \cdots, -1\cdot h_{1}, \cdots, 0\cdot h_{2}, \cdots .
\end{equation}

Moreover, applying eq. (\ref{eq:rcv_mod}) through a real system with
a block length $L = 1200$, the output of the decoder is depicted in
Fig. \ref{fig:llr_out_bpsk}. From the results shown in Fig.
\ref{fig:llr_out_bpsk} we can see clearly that the LLR contribution
of the second half of the systematic bits is marginal compared to
the first half. Each systematic half is under fadings $h_{1}$ and
$h_{2}$ respectively. Nevertheless, it was assumed that the
Root-Check codes would be able to recover all the systematic bits
under the same scenario discussed in this document.

Furthermore, we have investigated if we were properly generating a
block-fading channel using a complex channel. What instigated us to
conduct this analysis, it was based on the fact that from the
decoder point of view the received symbols must be in a binary phase
shift keying (BPSK) style. Consequently, we have obtained the
message LLR output shown in Fig. \ref{fig:llr_out_dec}.

The results presented in Fig. \ref{fig:llr_out_dec}, suggest that it
is not a block-fading $F = 2$ behaviour. However, if we just
multiply the channel matrix ($h_{1}$ or $h_{2}$) to its complex
conjugate to obtain the envelope of that channel realization, we
obtain a proper block-fading channel as it can be seen on Fig.
\ref{fig:llr_out_dec_opt}. We have noticed that the parity check
nodes of a general Root-Check LDPC codes tend to not converge, which
means in an IDD process the extrinsic information exchanged between
the decoder and the detector might not be fruitful in terms of
improving the overall performance. One interesting approach is the
controlled doping via high-order Root-Checks in graph codes
presented in \cite{doping.boutros.11}. The codes designed in
\cite{doping.boutros.11} are able to guarantee the parity check
nodes converge.

Following the analysis done for a SISO case we have now considered a
MIMO $2 \times 2$ signalling system. The channel can be interpreted
as a block-fading channel with $F = 2$, the LDPC code word has a
block length of $L = 1200$, maximum $20$ inner iterations and code
rate $R = \frac{1}{F} = \frac{1}{2}$. The Root-Check LDPC code used
is the one presented in \cite{doping.boutros.11}. The LLR output of
the decoder can be represented as:
\begin{equation} \label{eq:llr_output}
L_{C}(b_{t}) = \log \frac{P(b_{t} = +1\vert r_{t},\mathbf{H}_{f})}{P(b_{t} = -1\vert r_{t},\mathbf{H}_{f})}-L_{A}(b_{t})
\end{equation}
where $b_{t}$ is the t$-th$ bit and $L_{A}(b_{t})$ is the \textit{a priori} probability of the the t$-th$ bit. At the first outer iteration $L_{A}(b_{t}) = 0$. Furthermore, we will be looking at the \textit{a posteriori} LLR output of the decoder. For the case of SIC we have the symbols versus LLR output as shown in Fig. \ref{fig:llr_output_non_comp}. From the figure, we can see that the LLR of the parity bits (600 up to 1200) half of then are about $L_{C}(b_{t}) = \pm 120$ while the rest are even below $L_{C}(b_{t}) = \pm 80$. Therefore, we have an average gap of $40$ in terms of LLR.

We have adopted an LLR processing scheme for IDD with block-fading
channels (LLR-PS-BF). We have observed that before calculating the
extrinsic information for the next outer iteration, we calculate
\begin{equation} \label{eq:gamma}
\gamma = \max \|L_{C}^{'}(b_{L\times R}, \cdots, b_{L}) \|.
\end{equation} The reason why we have chosen the absolute maximum value of $L_{C}^{'}$ is whether we choose an arbitrary upper bound this leads to a degradation in the overall performance.

Accordingly, the final LLR $L_{C}$ with respect to the parity check
becomes:
\begin{equation} \label{eq:llr_output_mod}
L_{C}(b_{t}) = \mathrm{sgn}(L_{C}^{'}(b_{t}))\times \gamma ,~~~L\cdot R \leq t\leq L.
\end{equation}
It must be noted that in (\ref{eq:llr_output_mod}) every time the decoder generate an \textit{a posteriori} LLR $L_{C}^{'}$ a new value for $\gamma$ must be obtained. The main purpose of performing (\ref{eq:llr_output_mod}) is to enable a convergence for the detector in the next outer iteration. Therefore, the extrinsic message exchanged between the decoder and the detector will benefit from this operation. Consequently, a better performance in terms of BER will be noticed. Applying our proposed scheme LLR-PS-BF to LLR output obtained in Fig. \ref{fig:llr_output_non_comp} prior to the extrinsic information calculation, the final output is presented in Fig. \ref{fig:llr_output_comp}.

%%%%%%%%%%%%%%%%%%%%%%%%%%
%%%%%%%%%%%%%%%%%%%%%%%%%%

One can claim that it is not fair to only apply our method LLR-SP-BF to the Root-Check LDPC codes and not to apply for their counterpart, for instance, irregular repeat accumulate (IRA) LDPC codes. Correspondingly, we can show in Fig. \ref{fig:llr_out_non_comp_ira} that in the case of IRA LDPC code rate $R = \frac{1}{2}$, with block length $L = 1200$ there is no gap on the parity bits LLR. Moreover, we have applied our technique to see if this would bring some improvements. However, it has caused a degradation on overall performance. For the sake of clarity, all results that will be presented in Section \ref{sec:simul} based on IRA LDPC codes, we will not apply our method LLR-SP-BF to avoid a degradation on overall performance.

%%%%%%%%%%%%%%%%%%%%%%%%%%%%%%
The structure of our proposed LLR-PS-BF with soft cancellation detector is described in terms of iterations. In the first iteration, the \textit{a priori} information provided by the decoder is zero which heavily degrades the performance of PIC based detection. Therefore, we considered using PIC and SIC to see the overall performance. In the case of SIC approach, the soft estimates of $r_{t}$ is used to calculate the LLRs of their constituent bits. We assume $r_{t}$ is Gaussian, and, the soft output of the SISO detector for the $t$-th user is written as \cite{lamare.tcomms.08}
\begin{equation} \label{eq:soft_out_siso_dect}
\mathbf{r}_{t} = \mathbf{V}_{t} \mathbf{s}_{t}+\epsilon_{t},
\end{equation} where $\mathbf{V}_{t}$ is a scalar variable which is equal to the $t$-th users amplitude and $\epsilon_{t}$ is a Gaussian random variable with variance $\sigma_{\epsilon_{t}}^{2}$, with
\begin{equation} \label{eq:vk}
\mathbf{V}_{t} = E \left[\mathbf{s}_{t}^{*} r_{t}\right]
\end{equation} and
\begin{equation} \label{eq:sigma_ek}
\sigma_{\epsilon_{t}}^{2} = E \left[\vert r_{t}-\mathbf{V}_{t}\mathbf{s}_{t} \vert^{2} \right].
\end{equation}
The estimates of $\mathbf{\hat{V}}_{t}$ and $\hat{\sigma}_{\epsilon_{t}}^{2}$ can be obtained by time averages of the corresponding samples over the transmitted packet.

After the first iteration, the soft cancellation MMSE performs PIC and SIC by subtracting the soft replica of MAI components from the received vector as
\begin{equation} \label{eq:soft_cancelation}
\mathbf{\breve{r}}_{t} = \mathbf{r}_{t} - \mathbf{H}_{f} \mathbf{z}_{t},
\end{equation}
where $\mathbf{z}_{t} = \left[r_{1}, \cdots, r_{t\cdot \frac{N}{2}-1}, 0, r_{t\cdot \frac{N}{2}+1}, \cdots, r_{t\cdot N}\right]$ and a filter is developed to further reduce the residual interference as
\begin{equation} \label{eq:mmse_filter}
\mathbf{\omega}_{t} = \underset{\mathbf{\omega}_{t}}{\operatorname{arg~min}} E \lbrace \vert \mathbf{s}_{t} - \mathbf{\omega}_{t}^{H} \mathbf{\breve{r}}_{t}\vert^{2} \rbrace,
\end{equation}
where the soft output of the filter is also assumed Gaussian. The first and the second-order statistics of the symbols are also estimated via time averages of (\ref{eq:vk}) and (\ref{eq:sigma_ek}). The pseudo code of our proposed LLR-SP-BF IDD scheme is presented in Algorithm \ref{alg:llr_sp_bf_alg1}.

\begin{algorithm}
 \caption{Proposed LLR-SP-BF IDD Scheme}
 \label{alg:llr_sp_bf_alg1}
\algsetup{
linenosize=\small,
linenodelimiter=.
}
\begin{algorithmic}[1]
\STATE \textbf{Require}: $\mathbf{r}_{t} \in \chi_{t}^{\pm~1}$, $\mathbf{H}_{f} \in \chi^{n_{rx} \times n_{rt}}$, constellation set $\mathit{\chi}$, $\sigma_{v}^{2}$, $\mathbf{L}_{A}(b_{t})$ \textit{a priori} information, $\mathit{TI}$.
\medskip
\FOR{$\mathrm{l0} = 1 \to TI~\lbrace \mathrm{Turbo~Iteration}\rbrace$}
\medskip
    \STATE Calculate MMSE filter
    $\mathbf{W} = \left(\mathbf{H}_{f}\mathbf{H}_{f}^{H} + \frac{\sigma_{v}^{2}}{\sigma_{s}^{2}} \mathbf{I} \right)^{-1} \mathbf{H}_{f}$
\medskip
    \STATE \textbf{Detection Schemes} \\
    $\mathbf{r}_{\mathrm{SIC}} = \mathrm{Perform-SIC}(\mathbf{r}_{t}, \mathbf{H}_{f}, \sigma_{v}^{2}, \mathbf{W})$, perform the MMSE SIC detection scheme.\\
    $\mathbf{r}_{\mathrm{PIC}} = \mathrm{Perform-PIC}(\mathbf{r}_{t}, \mathbf{H}_{f}, \sigma_{v}^{2}, \mathbf{W})$, perform the MMSE PIC detection scheme.
\medskip
    \STATE \textbf{Obtain The Extrinsic Bit LLR}
\medskip
    \STATE \textbf{First}: Determine $\sigma_{\epsilon_{t}}^{2}$ based on the best channel realization by means of calculating:
    $\delta_{f} = \underset{f=1 \mathbf{to} F}{\operatorname{arg~max}}\vert \det(\mathbf{H}_{f})\vert$
\medskip
    \STATE $\delta_{f}$ is the index of $f$ which $\vert \det(\mathbf{H}_{f})\vert$ has the maximum value. Therefore, $\mathbf{V}_{t}$ and $\sigma_{\epsilon_{t}}^{2}$ must be calculated where the fading happens at index $\delta_{f}$. This is unique for block-fading channels, other types of channels do not require these additional steps. Then, the extrinsic LLR is obtained as: \\
    $L_{D}(b_{t})_{\mathrm{SIC}} = \log \frac{\sum_{\mathbf{s}_{t}~\in \chi_{t}^{+1}}P(\mathbf{r}_{\mathrm{SIC}}\vert s_{t}, \mathbf{H}_{f})P(\mathbf{s}_{t})}{\sum_{\mathbf{s}_{t}~\in \chi_{t}^{-1}}P(\mathbf{r}_{\mathrm{SIC}}\vert s_{t}, \mathbf{H}_{f})P(\mathbf{s}_{t})}~-~L_{C}(b_{t})$
    $L_{D}(b_{t})_{\mathrm{PIC}} = \log \frac{\sum_{\mathbf{s}_{t}~\in \chi_{t}^{+1}}P(\mathbf{r}_{\mathrm{PIC}}\vert s_{t}, \mathbf{H}_{f})P(\mathbf{s}_{t})}{\sum_{\mathbf{s}_{t}~\in \chi_{t}^{-1}}P(\mathbf{r}_{\mathrm{PIC}}\vert s_{t}, \mathbf{H}_{f})P(\mathbf{s}_{t})}~-~L_{C}(b_{t})$
\medskip
    \STATE \textbf{LDPC Decoding With Log-SP}
\medskip
    \STATE Obtain the a posteriori LLR of the decoder for both SIC and PIC, $L_{C}(b_{t})_{\mathrm{SIC}}$ and $L_{C}(b_{t})_{\mathrm{PIC}}$.
\medskip
    \IF{$\mathrm{LDPC} = \mathrm{RootCheck}$}
\medskip
        \STATE \textbf{Apply the proposed LLR-PS-BF}\\
        $\gamma = \max \|L_{C}^{'}(b_{L\times R}, \cdots, b_{L}) \|$,\\
        $L_{C}(b_{t}) = \mathrm{sgn}(L_{C}^{'}(b_{t}))\times \gamma ,~~~L\cdot R \leq t\leq L$;
\medskip
        \STATE Calculate the extrinsic information based on $L_{C}(b_{t})$ to be sent to the detector as \textit{a priori} information.
\medskip
    \ELSE
\medskip
        \STATE Directly calculate the extrinsic information based on $L_{C}(b_{t})$ to be sent to the detector as \textit{a priori} information.
\medskip
    \ENDIF
\medskip
\ENDFOR
\end{algorithmic}
\end{algorithm}

\section{Simulations} \label{sec:simul}
The bit error rate (BER) performance of the proposed LLR-PS-BF with PIC and SIC IDD scheme is compared with Root-Check LDPC codes and LDPC codes using a different number of antennas. Both LDPC codes used in the simulation are with block length $L = 1200$ for rate $R = \frac{1}{2}$. For rate $R = \frac{1}{4}$ the block length is $L = 1600$. The maximum number of inner iterations was set to $20$ and a maximum of $5$ outer iterations were used. We considered the proposed algorithms and all their counterparts in the independent and identically-distributed (i.i.d) random fast and block fading channels models. The coefficients are taken from complex Gaussian random variables with zero mean and unit variance. The modulation used is QPSK. The final SNR seen by the receiver is calculated as $SNR_{RCV} = \frac{1}{2\cdot \sigma_{\epsilon_{t}}^{2}}$ which is based on equation (\ref{eq:sigma_ek}).

\subsection{Single User Multiple Antennas}
\begin{figure}[htb]
 \centering
\resizebox{88mm}{!}{
\includegraphics{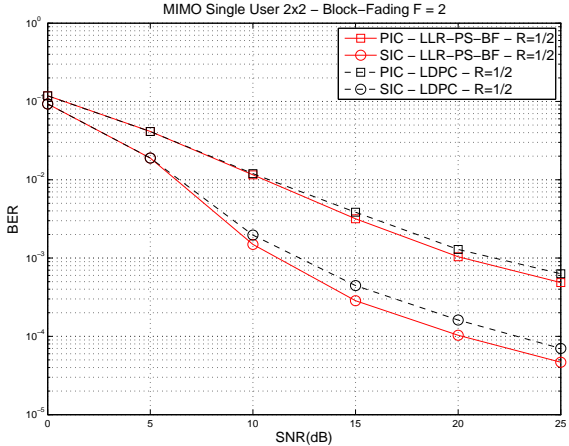}
}\caption{BER performance of LLR-PS-BF with Root-Check LDPC versus LDPC code both codes are rate $R = \frac{1}{2}$ and block length $L = 1200$. MIMO system with $2 \times 2$ antennas under block-fading channel with $F = 2$, QPSK modulation, $5$ outer iterations and $20$ inner iterations.} \label{fig:2x2subf2}
\end{figure}

% 2x2, RCDO VS IRA 3dB F = 2 R = 1/2
In Fig. \ref{fig:2x2subf2} the results for a MIMO system with $2 \times 2$, single user, block-fading channel with $F = 2$ fadings and LDPC codes rate $R = \frac{1}{2}$ are presented. We can see clearly for both SIC and PIC that our proposed scheme LLR-PS-BF with Root-Check LDPC codes outperforms the LDPC code by about $3dB$ in terms of SNR.

\begin{figure}[htb]
 \centering
\resizebox{88mm}{!}{
\includegraphics{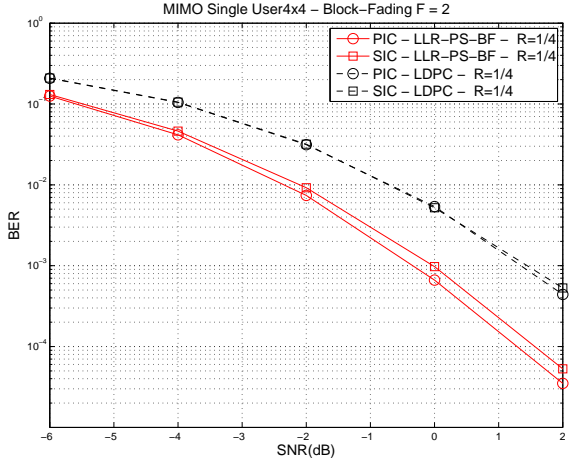}
}\caption{BER performance of LLR-PS-BF with Root-Check LDPC versus LDPC code both codes are rate $R = \frac{1}{4}$ and block length $L = 1600$. MIMO system with $4 \times 4$ antennas under block-fading channel with $F = 2$, QPSK modulation, $5$ outer iterations and $20$ inner iterations.} \label{fig:4x4subf2}
\end{figure}

% 4x4, RCDO VS IRA F = 2 R = 1/4
Fig. \ref{fig:4x4subf2} presents the results for a MIMO system with $4 \times 4$, single user, block-fading channel with $F = 2$ fadings and LDPC codes rate $R = \frac{1}{4}$. We can see clearly for both SIC and PIC that our proposed scheme LLR-PS-BF with Root-Check LDPC codes again outperforms the LDPC by about $2dB$ in terms of SNR. It can be noted that the SNR range presented in Fig. \ref{fig:4x4subf2} is much lower than the range in Fig. \ref{fig:2x2subf2}, this happens because in a MIMO system $4 \times 4$ the diversity order is $d_{max} = n_{rx}n_{tx} = 16$ as stated in \cite{mimobf.11}. Therefore, the upper bounded maximal code rate is $R = \frac{1}{n_{tx}} = \frac{1}{4}$.

\begin{figure}[htb]
 \centering
\resizebox{88mm}{!}{
\includegraphics{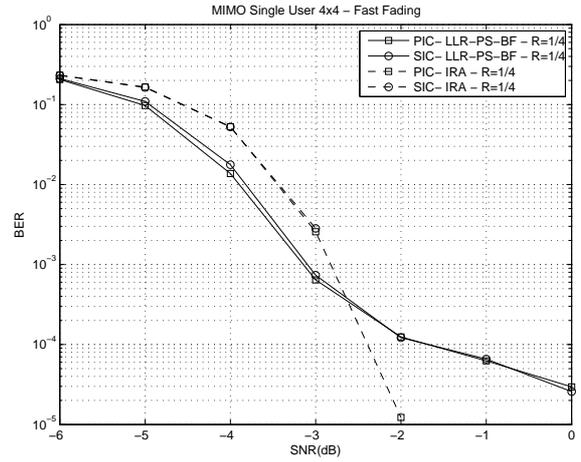}
}\caption{BER performance of LLR-PS-BF with Root-Check LDPC versus LDPC code both codes are rate $R = \frac{1}{2}$ and block length $L = 1200$. MIMO system with $2 \times 2$ antennas under fast fading channel, QPSK modulation, $5$ outer iterations and $20$ inner iterations.} \label{fig:2x2sufast}
\end{figure}

% 2x2, RCDO VS IRA 1dB FAST R = 1/2
Fig. \ref{fig:2x2sufast} shows the outcomes for a MIMO system with $2 \times 2$, single user, fast fading Rayleigh channel and LDPC codes rate $R = \frac{1}{2}$. In this channel the combination of Root-Check LDPC codes with our scheme LLR-PS-BF has provided a gain of $1dB$ with respect to the LDPC codes. In a fast fading channel the convergence for a lower BER is much earlier than in a block fading channel. Furthermore, in fast fading channel the coherence channel time is much less than symbol period $T_{C} \ll T_{S}$.

\section{Conclusion}
In this paper, we have presented a strategy based on the observation of the LLR output of the decoder for IDD systems. The proposed LLR-SP-BF algorithm has provided up to $3dB$ for a $2 \times 2$ MIMO and up to $2dB$ for a $4\times 4$ MIMO systems both over a block-fading channel with $F = 2$. For the case of a $2\times 2$ MIMO system over fast-fading the proposed LLR-SP-BF IDD scheme has obtained a gain of up to $1dB$. All the gains obtained in this paper are with respect to a MIMO IDD system using LDPC codes. There are some ongoing works to be done: first, to extend the proposed LLR-SP-BF strategy for Multi-User MIMO systems; second, to use scheduling techniques as proposed in \cite{schedul.wesel}; third, we are considering improved decoding strategies \cite{vfap.12} for the above mentioned strategies.

\section*{Acknowledgment}
This work was partially supported by CNPq (Brazil), under grant 237676/2012-5.

\bibliographystyle{IEEEtran}

%\bibliography{IEEEabrv,conf_paper}

\end{document}